\documentclass[preprint,superscriptaddress]{revtex4-1}
\usepackage{cmap}
\usepackage[utf8]{inputenc}
\usepackage[english]{babel}
\usepackage{amssymb,amsfonts,amsmath,mathtext,longtable,array}
\usepackage{xcolor}

\usepackage{graphicx}
\graphicspath{{1/}}

\usepackage[unicode,colorlinks]{hyperref}
\usepackage{breakurl}
\usepackage{soul}
\begin{document}
\UseRawInputEncoding

\title{Testing classicality of gravity by gravitation decoherence
}
\author{V.~Stefanov}
\affiliation{Institute of Applied Physics, University of Bern, Switzerland}

\author{D.~Mogilevtsev}
\affiliation{B.I.Stepanov Institute of Physics of NAS of Belarus, Belarus}

\author{I.~Rybak}
\affiliation{CAPA \& Departamento de F\'{\i}sica Te\'{o}rica, Universidad de Zaragoza,  Spain}
\affiliation{Instituto de Astrof\'{\i}sica e Ci\^encias do Espa\c co,
  CAUP, Portugal}
\affiliation{Centro de Astrofı́sica da Universidade do Porto, Portugal}

\author{A.~Stefanov}
\affiliation{Institute of Applied Physics, University of Bern, Switzerland}

\begin{abstract}
Here we discuss an influence of an external weak gravitational field on the gravitational self-decoherence effect with help of the stochastic extension of regularized Shr\"odinger-Newton equation in a curved background. We derive the master equation and demonstrate that it leads to the experimentally verifiable conclusions about applicability of the classical description of the weak gravitation field. Namely, a presence of oscillating terms in the otherwise purely exponential decay of the coherence would indicate classicality of gravity for the case. 
\end{abstract}

\maketitle

\section{Introduction}

Decoherence is a universal phenomenon that leads to the disappearance of quantum correlations and coherence. It is considered the primary reason for the observable classicality of our everyday world \cite{Zeh1970,Zurek}. In the quantum domain, decoherence typically occurs as a consequence of the interaction between a quantum system and its surroundings. For example, in solid-state systems such as quantum dots, interactions with phonons and structural defects can cause random shifts in transition frequencies, resulting in the decay of coherence on timescales several orders of magnitude faster than energy loss \cite{RevModPhys.76.1267}. 

To describe decoherence, several powerful and efficient open-system approaches have been developed and are now widely used, such as Langevin equations, master equations, and quantum trajectories \cite{breu}. Each approach has its own advantages and shortcomings, but all lead to the same predictions for observable results.

Gravitational decoherence presents a different situation. For over half a century, it has been recognized that gravitation, even by itself and through gravitational self-interaction, can act as a ``surrounding,'' inducing phase fluctuations and causing decoherence (see, for example, the review Ref.~\cite{Bassi_2017}). However, unlike well-understood quantum mechanics in flat space, even in the limit of weakly curved space, attempts to introduce gravitational effects into the quantum mechanical framework have led to ambiguous results.

First, destructive phase shifts similar to those caused by environmental influences can emerge from treating the gravitational field classically through time-dilation effects. A notable example Ref. \cite{Pik2015}, which explores the decoherence arising from the coupling between the internal degrees of freedom of a composite particle and its center of mass. This work has generated considerable discussion \cite{bonder2015,Pikovski_2017,Di_si_2017,PhysRevLett.117.090401}. Another relevant study investigates the gravitational disruption of directionality in the process of single-photon emission from an atomic ensemble~\cite{ourprd2020}.  

On the other hand, one can quantize gravity in a linear approximation, treat is as a reservoir, and derive master equations that describe dephasing in a fully quantum mechanical manner  \cite{PhysRevLett.111.021302}. Between these two approaches, there are several intermediate collapse models attempting to introduce dephasing noise in a more phenomenological way. Notable examples include the continuous spontaneous localization (CSL) model \cite{ghirardi1990continuous}, the Di\'osi-Penrose (DP) model \cite{diosi1989models, penrose1994shadows}, Adler’s proposal \cite{adler2001generalized} and Karolyhazy’s model \cite{karolyhazy1966gravitation}.

Thus, the question arises of possible experimental verification of the decoherence phenomenon, which could provide an opportunity to accept or reject particular methods of its description. {One should mention that recently quite a lot of attention was paid to test manifestations of quantum effects in curved state-time and the very manifestation of gravity quantumness. The reader can refer to the very recent comprehensive review \cite{RevModPhys.97.015003}. }

Here, we are addressing a particular question of necessity to use quantum description of gravity in a simple but methodologically important realistic scenario. We propose to perform testing of a quantum system decoherence, specifically by estimating the decay of the off-diagonal terms in the superposition of two different spatial positions of the same massive particle. We demonstrate how an experiment could be designed to differentiate between quantum and classical gravity models, specifically focusing on the stochastic extension of Schr\"odinger-Newton equation (SNe)~\cite{sne, nimmrichter2015stochastic}.

The SNe emerges when gravitational self-interaction terms are incorporated into the Shr\"odinger equation, introducing nonlinearity \cite{sne}. This form of gravitational self-interaction has been shown to account for the collapse of a pure wave function into a mixed state~\cite{DIOSI1984199, Penrose1996}. However, the derivation of the SNe is only valid under the assumption of fundamentally classical gravity~\cite{Anastopoulos_2014}. Therefore, this model serves as a potential test-bed for determining whether gravity is fundamentally quantum or classical, as discussed in Refs.~\cite{Carlip_2008, Giulini_2011, Grossardt, Oppenheim:2018igd, Oppenheim:2022xjr}.

Additionally, the conventional SNe is deterministic and allows for superluminal signaling~\cite{Gisin, Bassi_2013}. To prevent faster-than-light communication, the SNe can be regularized by introducing specific Brownian noise  \cite{nimmrichter2015stochastic, Bera}, leading to a linear master equation of the standard Lindblad form, with a collapse operator describing the decay of quantum coherence.

Previous studies of the self-interacting SNe have been conducted only in a Minkowski space background. Here, we extend this analysis to a weakly curved background. Starting with the SNe in an external weak gravitational field, we apply the regularization procedure from Ref.~\cite{nimmrichter2015stochastic} to derive the master equation in Lindblad form. The key point of our discussion is the fact  of qualitatively different master equation structures in the cases of quantum and semiclassical approaches for the the description of external gravitational field.   

Namely, the semi-classical approach in the small-time limit leads to the  modification of the exponential decay rate of coherence in comparison with the self-decoherence and appearance of a phase shift linearly growing with time.  
Quantumness of gravitational fields leads to the symmetrical quadratic form structure of the master equation precluding an appearance of the dynamics described above.  
Thus, experimental investigation of the position decoherence of a massive particle in an external gravitation field might indeed reveal an applicability of the semi-classical description of the weak gravitational field.

{The outline of the paper is as follows. 
For the sake of self-consistency, in Section \ref{Sec2} we provide a methodology for regularizing the SNe to address issues with superluminal signaling. In Section \ref{Sec3}, we extend this framework to include the presence of an external gravitational field.  In Section \ref{Sec5}, we evaluate the decoherence and phase shifting rates for a massive particle in an external gravitational field in the small-time limit. In Section \ref{Sec4} we describe the fundamental differences between quantum and classical consideration of the gravitational field considered as dephasing reservoirs and outline the difference in possible experimental outcomes. Finally, we conclude in Section \ref{Sec6}.

\section{Stochastic SNe master equation in the flat space-time background}
\label{Sec2}

In this section we briefly review derivation of the SNe describing gravitational self-action of a quantum particle in a flat space-time background to provide all necessary tools for further discussion of the SNe in a weakly curved space-time background. 

First of all, we assume that internal quantum degrees of freedom of the particle are not coupled with the particle position. Also, we describe gravitational self-interaction of the particle in the semi-classical manner\footnote{It is crucial to emphasize that \textit{semi-classical} in this context means that gravity is assumed to be fundamentally classical while the matter is quantum. See Ref.~\cite{bahrami2014schrodinger} for more details.}  writing the Einstein equations for the case in the following form \cite{KIEFER_book}:
\begin{equation}
\label{EinstEq}
R_{\mu\nu}+\frac{1}{2}g_{\mu\nu}R=\frac{8\pi G}{c^4}\langle \hat{T}_{\mu\nu}\rangle,
\end{equation}
where Greek indexes run over $4$ space-time coordinates ($0 \dots 3$), $R_{\mu \nu}$ is a Ricci tensor, $g_{\mu \nu}$ is a metric and $R = R^{\mu}_{\mu}$, while $\langle\ldots\rangle$ denote the averaging over the quantum state of the particle and   $\hat{T}_{\mu\nu}$ is the stress-energy operator. 

Considering self-acting gravitational field to be weak, we can linearize the Einstein's equations such that the space-time metric $g_{\mu \nu}$ can be written as   
\begin{equation}
\label{LineMink}
    g_{\mu\nu}=\eta_{\mu\nu}+h_{\mu\nu},
\end{equation}
where $\eta_{\mu \nu}$ is a Minkowski metric and $|h_{\mu \nu}|\ll1$ is a small perturbation \cite{Weinberg_book}. We impose harmonic coordinate conditions (de Donder gauge)
\begin{equation}
\label{DonderCond}
\partial^\mu \left(h_{\mu\nu}-\frac{1}{2}\eta_{\mu\nu}h^\alpha_\alpha\right)=0,
\end{equation}
where $\partial_{\mu} \equiv \partial/\partial x^{\mu}$.
Using conditions (\ref{DonderCond}), we write down the field equations (\ref{EinstEq}) in the following form
\begin{equation}\label{h_mn_vac}
    \square h_{\mu\nu}=\frac{16 \pi G}{c^4}\left(\frac{\eta_{\mu\nu}    \langle \eta^{\lambda \rho}\hat{T}_{\lambda \rho}\rangle}{2}-\langle  \hat{T}_{\mu\nu}\rangle\right),
\end{equation}
where $\square\equiv\partial_t^2-\Delta$ is a d'Alembert operator and $\Delta\equiv\partial_x^2+\partial_y^2+\partial_z^2$ is a Laplace operator in Minkowski space.

Now let us specify the perturbation in Eqs.~\eqref{LineMink} and \eqref{h_mn_vac} as being produced by the self-action of the considered particle. 
In the Newtonian limit $ T_{00}$ is significantly larger than the other nine stress-energy tensor components. So, the Hamiltonian describing interaction of the quantum particle and classical gravitation field in the Hamiltonian takes the following form \cite{adler2016gravitation}:
\begin{gather}
    \hat{H}_{\text{int}}=\frac{1}{2}\int d^3 r h_{\mu\nu} \hat{T}^{\mu\nu}\approx\frac{1}{2}\int d^3 r h_{00} \hat{T}^{00}=-G\int d^3 r\int d^3 r' \frac{\langle  \hat{\varrho}(\textbf{r}')\rangle}{|\textbf{r}-\textbf{r}'|}\hat{\varrho}(\textbf{r}),
    \label{hint}
\end{gather}
where the component $ \hat{T}^{00}=c^2\hat{\varrho}$ in the non-relativistic limit, with the mass density operator $\hat{\varrho}$. 

The Schr\"odinger equation with the total Hamiltonian $\hat H_0+\hat H_{int}$, where $\hat H_0=\frac{1}{2}\int d^3 r \eta_{\mu\nu} \hat{T}^{\mu\nu}$, is referred to as the  Schr\"odinger-Newton equation \cite{sne}. As mentioned in the Introduction, the SNe is believed to account for decoherence due to self-gravitation. However, SNe itself encounters several mathematical and physical challenges, notably violating the no-signaling condition \cite{Gisin, Bassi_2013}. 

It is possible to modify the SNe by introducing external noise and unravel it into well-defined master equation \cite{nimmrichter2015stochastic, Bera}.  The Hamiltonian (\ref{hint}) can be recast in a simple form exploiting the momentum representation of the mass density operator
\begin{equation}\label{hint2}
    \hat{H}_{\rm int}=\int d^3 \mathbf{k} \int d^3 \mathbf{k}' V(\mathbf{k},\mathbf{k}') 
    \langle {\hat {L}}_\mathbf{k}^\dagger\rangle  \hat L_{\mathbf{k}'}
\end{equation} 
 introducing the  operators  
\begin{eqnarray}
 \hat L_\textbf{k}=\int d^3 r \text{e}^{-i \textbf{k}\textbf{r}} \hat\varrho(\textbf{r}),
  \label{definition_L}
   \end{eqnarray}
and the kernel 
\begin{equation}
\label{kernel}
    V(\mathbf{k},\mathbf{k}')=-\frac{G}{(2\pi)^6}\int d^3 r\int d^3 r' \frac{\text{e}^{i( \textbf{k}'\textbf{r}'-\textbf{k}\textbf{r})}}{|\textbf{r}-\textbf{r}'|}
\end{equation}

The  form (\ref{hint2}) reminds one of coupling Hamiltonians typical for studies of open systems \cite{breu}.  The phenomenological approach suggested in Refs. \cite{nimmrichter2015stochastic, Bera} assumes introducing Lindblad operators in the following way
\begin{equation}
\hat{\operatorname L}_{\textbf{k}}= \hat L_{\textbf{k}}+i\langle \hat L_{\textbf{k}}\rangle
    \label{shiftedL}
\end{equation}
and associating with each operator $\hat{\operatorname L}_{\textbf{k}}$ an independent noise source described by the  Wiener stochastic increment  $dW_k$ with statistical averaging $\mathbb{E}[dW_k]=0$ and the following correlation  function \cite{DIOSI1984199}:
\begin{eqnarray}\label{cor1}
 \mathbb{E}[dW^*_{k'}dW_k]= \mathfrak{G}(\textbf{k}) \delta(\textbf{k}-\textbf{k}'), \\
    \mathfrak{G}(\textbf{k})=\frac{G}{2\hbar\pi^2 |\textbf{k}|^2}.  
\end{eqnarray}
Generally, the explicit form of $\mathfrak{G}(\textbf{k})$ depends on the collapse model.

By writing a stochastic diffusion It\^o equation for a quantum trajectory wave function and averaging it in the standard manner \cite{breu}, we obtain the following master equation for the density matrix $\rho$ of system
\begin{equation} \label{mas1}
    \dot{\rho}=-\frac{i}{\hbar} [H_0,\rho]+ \int d^3 k \mathfrak{G}(k)\left(\hat{\operatorname L}_\textbf{k} \rho \hat{\operatorname L}_{\textbf{k}}^\dagger-\frac{1}{2}\left\{\hat{\operatorname L}_\textbf{k}^\dagger \hat{\operatorname L}_{\textbf{k}}, \rho\right\}\right).
\end{equation}

Described by the standard Lindbladian master equation (\ref{mas1}), the density matrix $\rho$ is guaranteed to remain physical. However, one has to bear in mind that non-linearity is actually present in Eq.(\ref{mas1}), despite the absence of a nonlinear Hamiltonian self-interaction term. 

\section{Stochastic extension of SNe in an external weak gravitational field}
\label{Sec3}

Here, we derive the SNe master equation for a curved background, specifically considering the Schwarzschild metric, $g^{\rm (SC)}_{\mu\nu}$, 
using isotropic coordinates $\{t,x,y,z\}$ in the following way \cite{peters1966perturbations}: 
\begin{eqnarray}
\label{metric}
   &&g_{00}^{\rm (SC)}=\left( \frac{1+\phi/2}{1-\phi/2} \right)^2 ,\\
    &&g_{0i}^{\rm (SC)}=0,\\
    &&g_{ij}^{\rm (SC)}=-\delta_{ij} \left( 1-\phi/2 \right)^4,
\end{eqnarray}
where Latin indexes run over only spatial coordinates (from 1 to 3), $\phi=-r_s/2r$, $r=\sqrt{x^2+y^2+z^2}$, $\delta_{ij}$ is the Kronecker symbol, and $r_s=2GM/c^2$ is the Schwarzschild radius associated with the mass $M$.

As in the previous section, we treat the gravitational self-interaction of a particle as a small perturbation, expressing the resulting metric as follows:
\begin{equation}
    g_{\mu\nu}=g^{\rm (SC)}_{\mu\nu}+h_{\mu\nu},
\end{equation}
We also assume a weakly curved background, retaining only the terms that are linear in $\phi$ and its derivatives. 

Hence, we expand perturbation as $h_{\mu\nu}=h^{(0)}_{\mu\nu}+h^{(1)}_{\mu\nu}$, where $h^{(0)}_{\mu\nu}$ are perturbations on a flat background, while $h^{(1)}_{\mu\nu}$ are linear corrections due to background curvature.
From requirements that $\nabla^{\mu} \hat{T}_{\mu\nu}=0$, where $\nabla^{\mu}$ defines a covariant derivative, and $\int d^3 r \sqrt{-g} \langle \hat{T}^{0}{}_0\rangle=mc^2$ at $t=0$ \cite{gorbatsievich1985quantum} - with $\sqrt{-g}$ being the determinant of the metric (\ref{metric}) - one can derive the following expressions for $h^{(0)}_{00}$ and $h^{(1)}_{00}$ in the small-time limit (see Appendix \ref{AppA} for details)
\begin{gather} \label{Solution1}
  h^{(0)}_{00}(\textbf{r})=-\frac{2 G}{c^2}\int d^3 r' \frac{\langle \hat \varrho(\textbf{r}')\rangle}{|\textbf{r}-\textbf{r}'|},\\
   h^{(1)}_{00}(\textbf{r})=\frac{2G}{c^2}r_s\Bigg(\frac{1}{|\textbf{r}|}\int d^3 r'\frac{\langle \hat \varrho(\textbf{r}')\rangle}{|\textbf{r}-\textbf{r}'|}+\frac{1}{8\pi}\int d^3 r' \int d^3 r'' \frac{1}{|\textbf{r}-\textbf{r}'||\textbf{r}'-\textbf{r}''||\textbf{r}''|^2}\frac{\partial \langle \hat \varrho(\textbf{r}'')\rangle}{\partial|\textbf{r}''|}-\nonumber
   \\-\frac{(c t)^2}{2}\frac{1}{2}\int d^3 r' \frac{1}{|\textbf{r}-\textbf{r}'||\textbf{r}'|^2}\frac{\partial \langle \hat \varrho(\textbf{r}')\rangle}{\partial|\textbf{r}'|} \Bigg) \label{Solution2}
  \end{gather}

Performing Fourier transformation and replacing $\hat L_\textbf{k}$ according to Eq.~(\ref{definition_L}), we obtain interaction Hamiltonian $\hat{H}_{\rm int}$ in the form (\ref{hint2}) with the following kernel 
\begin{equation}
\label{kernel2}
    V(\mathbf{k},\mathbf{k}')=-\frac{G}{2\pi^2}\left(\alpha_{\textbf{k},\textbf{k}'}+\beta_{\textbf{k},\textbf{k}'}\right),
\end{equation} 
where symmetric term  $\alpha_{\textbf{k},\textbf{k}'}$ and anti-symmetric term $\beta_{\textbf{k},\textbf{k}'}$ are given by
\begin{equation}
\alpha_{\textbf{k},\textbf{k}'}\equiv\frac{F(\textbf{k}',\textbf{k})+F(\textbf{k},\textbf{k}')}{2},
\end{equation}
\begin{equation}
\label{beta}
\beta_{\textbf{k},\textbf{k}'}\equiv\frac{F(\textbf{k}',\textbf{k})-F(\textbf{k},\textbf{k}')}{2},
\end{equation}
with 

\begin{multline}\label{F_function}
 F(\textbf{k}',\textbf{k})\equiv\frac{1}{(2\pi)^3}\frac{\delta^3(\textbf{k}' -\textbf{k})}{|\textbf{k}'||\textbf{k}|}-\\-\frac{r_s}{4\pi^2}\frac{1}{ |\textbf{k}-\textbf{k}'|^2}\Bigg(\frac{\textbf{k} \cdot (\textbf{k}'-\textbf{k})}{k'^4}-\frac{(c t)^2}{2}\left(\frac{\textbf{k} \cdot (\textbf{k}'-\textbf{k})}{\textbf{k}'^2}+\frac{\textbf{k}' \cdot (\textbf{k}-\textbf{k}')}{\textbf{k}^2}\right)\Bigg) 
\end{multline}

The structure of Eq.(\ref{kernel2}) is the main key for our conclusions about applicability of the semi-classical description of the gravitational field. Presence of both the symmetric and antisymmetric terms points to the contribution of $\hat{H}_{\rm int}$ not only to dissipative, but also to unitary dynamics of the superposition state of our massive particle. Notice that this phenomenon arises solely due to presence of an external gravitational field. Self-gravitational interaction (i.e., for $r_s=0$) does not lead to this effect.

With It\^o calculus the diffusive extension of SNe has the same form as for vacuum case. The noise $dW_1$  the same properties: $\mathbb{E}[dW_1]=0$ and 
\begin{equation}
 \mathbb{E}[dW_1^*(\textbf{k}')dW_1(\textbf{k})] \equiv\mathfrak{G}(\textbf{k}',\textbf{k}),
\end{equation}
where correlation function explicitly depends on $\textbf{k}$ and $\textbf{k}'$ because of broken translation and rotation symmetries by the external gravitational field. 
Because of the hermiticity of the master equation, the correlation function must have  symmetrical properties under swapping $\textbf{k}'$ and $\textbf{k}$: $\mathfrak{G}(\textbf{k}',\textbf{k})^*=\mathfrak{G}(\textbf{k},\textbf{k}')$. Supposing $\mathfrak{G}(\textbf{k}',\textbf{k})=\alpha_{\textbf{k},\textbf{k}'}$ and using the same shift for operator: $\hat L_\textbf{k} \rightarrow \hat L_\textbf{k}+ i \langle \hat L_\textbf{k}  \rangle$, we can
eliminate the non-linear term by $\alpha_{\textbf{k},\textbf{k}'}(t=0)$ in the interaction Hamiltonian. 

To absorb the term with $\beta_{\textbf{k},\textbf{k}'}$ we should add another Wiener noise $dW_2$ with $\mathbb{E}[dW_2]=0$ and with correlation function
\begin{equation}
\mathbb{E}[dW_2^*(\textbf{k}')dW_2(\textbf{k})]\equiv i \beta_{\textbf{k},\textbf{k}'}.
\end{equation}
Performing the shift of operator: $\hat L_\textbf{k} \rightarrow \hat L_\textbf{k}- \langle \hat L_\textbf{k}\rangle$, the final form of master equation becomes 
\begin{equation} \label{master_equation_clas}
    \dot{\rho}=-\frac{i}{\hbar} [\hat{H}_0,\rho]+\frac{G}{2\pi^2\hbar}\int d^3 k\int d^3 k' \left(\alpha_{\textbf{k},\textbf{k}'}+i\beta_{\textbf{k},\textbf{k}'}\right)\times \left(L_{\textbf{k}} \rho L_{\textbf{k}'}^\dagger-\frac{1}{2}\left\{L_{\textbf{k}'}^\dagger L_{\textbf{k}}, \rho\right\}\right).
\end{equation}
Details about its derivation can be found in the Appendix \ref{AppB}.

It is impossible to eliminate non-linear terms in the interaction Hamiltonian of the SNe by one complex correlation function. We need to include two independent Wiener noises: $dW_1$ and $dW_2$. The origin of the $dW_1$ is a self-interaction of the local mass density that preserves rotation symmetry and leads to decoherence. While the noise $dW_2$ appears due to a weak external gravitational field, it breaks the spherical symmetry and induces a phase shifting effect. 

\section{Estimation of the correction to the decoherence rate}
\label{Sec5}

Let us now estimate the change of gravitational decoherence due to the external gravitational field.
For this, we write the Lindblad part of the master equation \eqref{master_equation_clas} in coordinate representation:
\begin{multline}\label{lindblad_coordiante_representation}
    \langle \textbf{x}|L(\rho)|\textbf{x}'\rangle=\frac{G{m^2}}{2\pi^2\hbar}\int d^3 k\int d^3 k' \left(\alpha_{\textbf{k},\textbf{k}'}+i\beta_{\textbf{k},\textbf{k}'}\right) \times\\ \times \left( \text{e}^{i\textbf{k}\textbf{x}-i\textbf{k}'\textbf{x}'}-\frac{\text{e}^{i(\textbf{k}-\textbf{k}')\textbf{x}}}{2} -\frac{\text{e}^{i(\textbf{k}-\textbf{k}')\textbf{x}'}}{2} \right)\langle\textbf{x}|\rho|\textbf{x}'\rangle,
\end{multline}
{where $L_\textbf{k}=m e^{i \textbf{k} \textbf{x}}$ (without cut-off). }
We consider that effective radius $\textbf{r}$ (space distribution) of a quantum object much smaller than the distance to the center of a massive body $\textbf{r}_0$: $\textbf{x}=\textbf{r}_0+\textbf{r}$, $\textbf{r}_0\gg \textbf{r}$ and expand the correction to the linear order of $\textbf{r}$ and $\textbf{r}'$ in the following expressions {(see details in Appendix \ref{AppC})}:
\begin{equation}\label{non_diag_cor_coh}
   \frac{G{m^2}}{2\pi^2\hbar}\int d^3 k\int d^3 k' \alpha_{\textbf{k},\textbf{k}'}\text{e}^{i\textbf{k}\textbf{x}-i\textbf{k}'\textbf{x}'}\approx   \frac{G{m^2}}{\hbar}\frac{1}{|\textbf{r}-\textbf{r}'|} \left(1-r_s\frac{|\textbf{r}-\textbf{r}'|^2}{8r_0^3}\left(1-3 \cos^2 \Gamma\right)\right),
\end{equation}
with $\Gamma$ is an angle between vectors $\textbf{r}_0$ and $\textbf{r}-\textbf{r}'$, $\cos \Gamma\equiv(\textbf{r}-\textbf{r}')\textbf{r}_0/ |\textbf{r}-\textbf{r}'| |\textbf{r}_0|$, and 
\begin{equation}
    \frac{G{m^2}}{2\pi^2\hbar}\int d^3 k\int d^3 k' \beta_{\textbf{k},\textbf{k}'}\text{e}^{i\textbf{k}\textbf{x}-i\textbf{k}'\textbf{x}'}\approx  - \frac{G{m^2}}{\hbar} r_s\frac{\cos \Gamma}{4r_0^2}.
\end{equation}
We get a divergence in \eqref{non_diag_cor_coh}, when $r\rightarrow r'$ due to the absence of a cut-off. To address this, we assume a coarse-grained mass density operator with a spatial resolution $R$ \cite{DIOSI1987377}.

So, if we neglect the pure Schr\"odinger contribution we can write down
\begin{gather}
    \langle \textbf{x} |\rho(t)| \textbf{x}'\rangle=\exp\left\{-\left(\Lambda^{\rm (dec)} + i \Lambda^{\rm (ph)} \right)t\right\}  \langle \textbf{x} |\rho(0)| \textbf{x}'\rangle,\nonumber\\
   \Lambda^{\rm (dec)}=\Lambda^{\rm (dec)}_{\rm flat}-\frac{G m^2}{\hbar}r_s\frac{|\textbf{r}-\textbf{r}'|}{8r_0^3}\left(1-3 \cos^2 \Gamma\right), \label{dec_rate}\\
    \Lambda^{\rm (dec)}_{\rm flat}=\frac{G m^2}{\hbar}\left(\frac{1}{\sqrt{\pi}R}-\frac{\text{Erf}\left(|\textbf{r}-\textbf{r}'|/2R\right)}{|\textbf{r}-\textbf{r}'|}\right),\nonumber\\
    \Lambda^{\rm (ph)}=\frac{G m^2 }{\hbar}r_s\frac{\cos \Gamma}{4r_0^2},\label{deph_rate}
\end{gather}
where $\Lambda^{\rm (dec)}_{\rm flat}$ defines the well-known decoherence rate for a quantum object in vacuum \cite{Bassi_2017} with Gauss error function $\text{Erf}(r)$.

{As we noted above, the dephasing $ \eqref{deph_rate} $ appears due to the external gravity field. If this field is so weak as it is at the Earth's surface, it is natural to expect an extreme weakness of the dephasing effects for such objects as single atoms or ions commonly taken as a testbed for quantum tests. However, the problem is mitigated by the current experimental possibilities of preparing coherent superpositions of massive particles. For example, current experimental technology allows creating Bose-Einstein condensate of rather large number (say, $10^9$ \cite{PhysRevLett.81.3811}) of heavy atoms, such as Rubidium, Dysprosium and Thulium \cite{PhysRevLett.108.210401,PhysRevLett.107.190401,PhysRevA.102.011302}. Thus at the Earth's surface, it is feasible getting $ \Lambda^{\rm (ph)}\approx 10^{-21}-10^{-22}$.  It is still a small value. However, the rate $ \Lambda^{\rm (ph)} $ does not depend on the cut-off or the distance between measurement points, but only on their orientation relative to the center of a massive body.
This feature of phase shifting is analogous to the Off-Diagonal Long-Range Order \cite{yang1962concept}, where the off-diagonal terms of the density matrix do not depend on the distance between the considered spatial points (which allows the detection of relative phase). Generally, we are discussing a phase effect, which, in principle, allows for very precise interferometric measurements. The most important point is that its detection is fundamentally possible, and experimental probes will advance our understanding of the nature of gravitational theory.
}

\section{Quantum description of the gravitational field}
\label{Sec4}

{Here, we aim to describe positional decoherence in a weak external gravitational field, considering gravity as fundamentally quantum. To achieve this, the metric perturbations $h_{\mu \nu}$ are treated as linear operators $\hat{h}_{\mu \nu}$ \cite{PhysRevD.50.3874}. By repeating the steps as in Appendix~\ref{AppA}, one can demonstrate that the equations for leading contribution have a form similar to the classical description (compare with \eqref{h00}-\eqref{h001}})\footnote{{We note that quantum radiative corrections in the weak-field approximation are of the order $ \sim G \hbar / c^3 r^2 $ \cite{bjerrum2003quantum}. For a laboratory on the Earth's surface, these corrections are negligable, because they have the order $\sim 10^{-83} $.}}:
{
\begin{eqnarray}
\label{QuantH0}
    &&\Delta \hat{h}^{(0)}_{00}=\frac{8\pi G}{c^4}\hat{T}_{00}^{(0)},\\
\label{QuantH1}
     &&\square \hat{h}^{(1)}_{00}+2\Delta (\phi \hat{h}^{(0)}_{00})-4 \phi \Delta  \hat{h}^{(0)}_{00}=-\frac{8\pi G}{c^4}\hat{T}_{00}^{(1)}.
\end{eqnarray}
}

{By explicitly solving for $\hat{h}^{(0)}_{00}$ and $\hat{h}^{(1)}_{00}$, similar to Eqs.~\eqref{Solution1}-\eqref{Solution2}, one can derive the interaction Hamiltonian in the following form:} 
\begin{equation}\label{H_int_quant}
    \hat{H}_{\rm int}=-\frac{G}{2\pi^2}\int d^3 k \int d^3 k' L_{\textbf{k}'}  L_\textbf{k}^\dagger  {\bar{V}}(\textbf{k},\textbf{k}'),
\end{equation} 
where the scalar kernel function  ${\bar{V}}(\textbf{k},\textbf{k}')$ is necessarily symmetric with respect to $\textbf{k}$ and $\textbf{k}'$. 

{The following transformation of the Shr\"odinger equation, considering the Hamiltonian \eqref{H_int_quant} allows to receive the SNe as a mean-field limit of many-particle system \cite{erdHos2001derivation,bardos2002derivation}, that is similar to classical case \eqref{master_equation_clas}, but with a few important differences. For the classical case of gravity, the interaction Hamiltonian \eqref{hint2} is linear with respect to the operators ${L}_{\textbf{k}}$ (or ${L}_{\textbf{k}}^{\dagger}$), while the 
fully quantum interaction Hamiltonian $(\ref{H_int_quant})$ is a symmetric quadratic form with respect to the operators   $L^{\dagger}_{\textbf{k}}$ and  $L_{\textbf{k}}$.}
This Hamiltonian gives no contribution to the unitary dynamics, and cannot lead to the linearly increasing phase-shift as an effect complimentary to the exponential dephasing coherence decay described by Eqs.~\eqref{dec_rate}-\eqref{deph_rate}. 

{
It is important to highlight that the prediction of gravitational decoherence in a flat space-time is actually the same for both classical and quantum approaches. However, in the case of an external gravity field,  predictions of the gravitational collapse for classical and quantum approaches are drastically different.  Thus, appearance of the dynamics accurately described by an exponential coherence decay with a linearly increasing  phase shift induced by an external gravitational field can serve as a test to distinguish whether weak gravitational fields allow for a classical description in an experimentally feasible scenario. There is no way to get it for the quantum approach.
}

\section{Conclusions}
\label{Sec6}

In this work we estimated a perturbation induced to a decoherence of a quantum superposition state  of a massive particle by a weak external spherically symmetric gravitational field. We demonstrated that this problem has a significance going far beyond a technical add-on to the corpus of existing works of gravitational decoherence. Namely, it supplies one with a tool for making  conclusions about quantumness or classicality of gravity.

The question of whether gravity is classical or quantum in nature remains open and continues to attract the attention of researchers \cite{Oppenheim:2018igd, Oppenheim:2022xjr, Alonso:2023ien, Tilloy:2024qpi}. From this perspective, the quantum approach to describing gravity could be incorrect, or at least requiring some re-thinking of the quantum theory foundations. This is an area that certainly requires further examination. Experimental validation of the predicted dynamics would allow one to certify (or at least to have a strong point in favor of) the classical nature of gravity. 

The very possibility of these conclusions stems from uncovered structural differences in the description of the quantum particle interaction with an external gravitational field. For a semi-classical description  a presence of an external gravitational field induces corrections to the dynamics leading to the phase-shift of off-diagonal terms of the density matrix  linearly increasing with time.  In contrast to it, the quantum description leads to the symmetric quadratic form of the interaction Hamiltonian precluding it from giving a contribution to the unitary part of the dynamics. So, registering an exponential coherence decay accompanied by a phase-shift would indicate an applicability of the semi-classical description of the gravitational field for the case. 

{Additionally, it is likely that this effect could much more observable near astronomical objects with stronger gravitational fields or could have astrophysical applications, particularly in models of ultralight dark matter particles \cite{Hui}. These particles are of particular interest because their quantum properties can persist for extended periods due to their low decoherence \cite{Allali:2021puy, Allali_2020, Eberhardt:2023axk, Cheong:2024ose}. In contrast, most other objects in the universe undergo rapid decoherence, quickly transitioning to classical behavior. However, the minimal decoherence of ultralight dark matter particles also makes it significantly more important for considering, although challenging to detect. However, their ability to maintain quantum coherence for a long time has made them a focal point in the search for potential quantum effects in cosmology \cite{Marsh:2022gnf}.}

\section{Acknowledgments}
D.M. acknowledges support from BRFFR grant F23UZB-064.

I.R. also acknowledges support from the Grant PGC2022-126078NB-C21 funded by MCIN/AEI/ 10.13039/501100011033 and ``ERDF A way of making Europe'', as well as Grant DGA-FSE grant 2020-E21-17R from the Aragon Government and the European Union - NextGenerationEU Recovery and Resilience Program on `Astrof\'{\i}sica y F\'{\i}sica de Altas Energ\'{\i}as' CEFCA-CAPA-ITAINNOVA.

\begin{widetext}
\appendix

\section{The Schr\"odinger-Newton equation in weak external field} \label{AppA}

Let find perturbations in the Schwarzchild metric $h_{\mu\nu}$ generated by some quantum object with TEM $\hat T_{\mu\nu}$:
\begin{equation}
    g_{\mu\nu}=g^{(S)}_{\mu\nu}+h_{\mu\nu}.
\end{equation}
Following \cite{peters1966perturbations} the Einstein equations have form:
\begin{multline}\label{peters_eqs}
\overline{h}_{\mu\nu;\alpha} {}^{;\alpha}-f_{\mu;\nu}-f_{\nu;\mu}+g^{(S)}_{\mu\nu}f_{\alpha}{}^{;\alpha}-\left(-2\overline{h}_{\alpha\beta}R^{\alpha}{}_{\mu\nu}{}^{\beta}+\overline{h}_{\mu\alpha}R^{\alpha}{}_{\nu}+\overline{h}_{\nu\alpha}R^{\alpha}{}_{\mu}-h_{\mu\nu}R+g^{(S)}_{\mu\nu}h_{\alpha\beta}R^{\alpha\beta}\right)=\\=-\frac{16\pi G}{c^4}\langle\hat T_{\mu\nu}\rangle,
\end{multline}
where $\overline{h}_{\mu\nu}=h_{\mu\nu}-1/2g^{(S)}_{\mu\nu}h_{\alpha}{}^{\alpha}$ and $f_{\mu} \equiv \overline{h}_{\mu\nu}{}^{;\nu}$ is an arbitrary vector function representing the gauge condition. The Ricci scalar is $R=R^\alpha{}_\alpha$, the Ricci curvature tensor is  $R_{\alpha\beta}=R^{\mu}{}_{\alpha\mu\beta}$ and Riemann curvature tensor $R^{\alpha}{}_{\mu\beta\nu}$ is derived with Christoffel symbols $\Gamma^\alpha{}_{\mu\nu}$ as following
\begin{eqnarray}\label{riemann}
R^{\alpha}{}_{\mu\beta\nu}=\partial_\beta \Gamma^\alpha{}_{\mu\nu}-\partial_\nu \Gamma^\alpha{}_{\mu\beta}+\Gamma^\alpha{}_{\beta\lambda}\Gamma^\lambda{}_{\mu\nu}-\Gamma^\alpha{}_{\nu\lambda}\Gamma^\lambda{}_{\mu\beta},\\
\Gamma^\alpha{}_{\mu\nu}=\frac{1}{2}g^{(S) \; \alpha\lambda}\left(\partial_\mu g^{(S)}_{\lambda\nu}+\partial_\nu g^{(S)}_{\lambda\mu}-\partial_\lambda g^{(S)}_{\mu\nu}\right).
\end{eqnarray}
Note, that we have used the definition \eqref{riemann} for the Riemann curvature tensor according to modern convention, and it differs in sign from the original one in \cite{peters1966perturbations}. As follows, we also changed sign before bracket in \eqref{peters_eqs}.

In isotropic coordinates $\{t,x,y,z\}$ with neglecting of powers of $\phi^2$ we have:
\begin{eqnarray}\label{metric_I}
g^{(S)}_{00}=(1+2\phi),g^{(S)}_{0i}=0,g^{(S)}_{ij}=-\delta_{ij}(1-2\phi),
\end{eqnarray}
with potential $\phi=-GM/rc^2$, $r=\sqrt{x^2+y^2+z^2}$, and $M$ is the mass of Schwarzchild body.

From energy conservation law
\begin{eqnarray}\label{eq:energy_cons_law}
\hat T_{\mu\nu}{}^{;\nu}=0,
\end{eqnarray}
and initial conditions (see section 3.5 of Ref.~\cite{gorbatsievich1985quantum}):
\begin{eqnarray}\begin{aligned}
E=m c^2=\int d\mathfrak{f}_\mu {\epsilon}^{\nu}\langle\hat T_{\nu}{}^\mu\{t=0\}\rangle=\int_{t=const}d^3 x n_\mu{\epsilon}^{\nu}\langle\hat T_{\nu}{}^\mu\{t=0\}\rangle=\int d^3 r  \sqrt{-g} \langle\hat T_{0}{}^0\{t=0\}\rangle,\\ 
 \hat T^{0i}\{t=0\}=0,\\
 \hat T^{ij}\{t=0\}=0,
\end{aligned}\end{eqnarray}
where ${\epsilon}^{\nu}=\{c,0,0,0\}$ is the timelike Killing vector of
our space-time \eqref{metric_I}, $\mathfrak{f}_\mu$ is a hypersurface, and $n_\mu=\{\sqrt{-g}/c,0,0,0\}$ is the orthogonal vector to that hypersurface; $g$ is a determinant of metric $g^{(S)}_{\mu \nu}$, 
we can find $\hat T_{\mu\nu}=\hat T^{(0)}_{\mu\nu}+\hat T^{(1)}_{\mu\nu}$ as sum of non-perturbing $\hat T^{(0)}_{\mu\nu}$ and linear  to $\phi$ perturbation $\hat T^{(1)}_{\mu\nu}$ terms:
\begin{eqnarray}\label{T_approx}\begin{aligned}
  \hat T^{(0)}_{00}=\hat\varrho c^2, \qquad \hat T^{(0)}_{0i}=0, \qquad \hat T^{(0)}_{ij}=0, \\
  \hat T^{(1)}_{00}=\left(4\phi\hat\varrho+\delta^{ij}\phi_{,i} \hat\varrho_{,j} \frac{(ct)^2}{2}\right) c^2, \qquad 
  \hat T^{(1)}_{0i}=\phi_{,i} \hat\varrho t c^2, \qquad  \hat T^{(1)}_{ij}=0,
\end{aligned}\end{eqnarray}
where $\langle\hat\varrho\rangle$ is average mass density ($\int \langle\hat\varrho\rangle d^3 r=m$).

Important comment about \eqref{eq:energy_cons_law}, its explicit form is a heat equation
, and it can be solved analytically in case of defining the initial and boarder conditions. The correspondent heat kernel is 
\begin{equation}
    \Phi\left(r,t\right)\sim \exp\left\{-\frac{(ct)^2}{4 \int\limits  \frac{dr}{\partial_r(\phi+\phi^2)}}\right\}\approx \exp\left\{-\frac{3r_s(ct)^2}{8r^3}\right\},
\end{equation}
with respect to linear terms to $\phi$, and where $r_s=2GM/c^2$ is a Schwarzschild radius. Therefore full solution of $\hat{T}_{\mu\nu}$ converges, and a range of variable $t$ for applicability of \eqref{T_approx} should be estimated from $3r_s(ct)^2/8r^3\ll 1$.

Taking the linear term to $\phi$ for $h_{\mu\nu}=h^{(0)}_{\mu\nu}+h^{(1)}_{\mu\nu}$ we can write in de Donder gauge ($f_\mu=0$) the following equations:
\begin{eqnarray}
 &\Delta h^{(0)}_{00}=\frac{8\pi G}{c^2}\langle\hat\varrho\rangle,\label{h00}\\
 &\square h^{(1)}_{00}+2\Delta (\phi h^{(0)}_{00})-4 \phi \Delta  h^{(0)}_{00}=-\frac{8\pi G}{c^2}\left(4\phi\langle\hat\varrho\rangle+\delta^{ij}\phi_{,i} \langle\hat \varrho\rangle_{,j} \frac{(ct)^2}{2}\right),\label{h001}
\end{eqnarray}
with  d'Alembert operator $\square\equiv\partial_t^2-\Delta$ and $\Delta\equiv\partial_x^2+\partial_y^2+\partial_z^2$. Separate time dependent part of $h^{(1)}_{00}$ as $h^{(1)}_{00}=q+w (ct)^2/2 $ we found:
\begin{eqnarray}
  \Delta w=\frac{8\pi G}{c^2}\delta^{ij}\phi_{,i} \langle\hat \varrho\rangle_{,j} ,\\
  \Delta \left(q-2\phi h^{(0)}_{00}\right)=w.
\end{eqnarray}
Here we neglect the time dependence of $\langle \varrho\rangle$ assuming that averaging value changes much slower then metric changes.

The final solution has form:
\begin{eqnarray}
  h^{(0)}_{00}(\textbf{r})=-\frac{2 G}{c^2}\int d^3 r' \frac{\langle \hat \varrho(\textbf{r}')\rangle}{|\textbf{r}-\textbf{r}'|},\\
  w(\textbf{r})=-\frac{2 G}{c^2}r_s\int d^3 r' \frac{1}{2|\textbf{r}-\textbf{r}'||\textbf{r}'|^2}\frac{\partial}{\partial|\textbf{r}'|}\langle \hat \varrho(\textbf{r}')\rangle,\\
  q(\textbf{r})=\frac{2 G}{c^2}r_s\frac{1}{|\textbf{r}|}\int d^3 r' \frac{\langle \hat \varrho(\textbf{r}')\rangle}{|\textbf{r}-\textbf{r}'|}-\frac{1}{4\pi}\int d^3 r' \frac{w(\textbf{r}')}{|\textbf{r}-\textbf{r}'|},
  \end{eqnarray}
and after combining
  \begin{multline}
  h^{(1)}_{00}(\textbf{r})=\frac{2G}{c^2}r_s\Bigg(\int d^3 r'\frac{1}{|\textbf{r}|} \frac{\langle \hat \varrho(\textbf{r}')\rangle}{|\textbf{r}-\textbf{r}'|}+\frac{1}{4\pi}\frac{1}{2}\int d^3 r' \int d^3 r'' \frac{1}{|\textbf{r}-\textbf{r}'||\textbf{r}'-\textbf{r}''||\textbf{r}''|^2}\frac{\partial}{\partial|\textbf{r}''|}\langle \hat \varrho(\textbf{r}'')\rangle-\\-\frac{(ct)^2}{2}\frac{1}{2}\int d^3 r' \frac{1}{|\textbf{r}-\textbf{r}'||\textbf{r}'|^2}\frac{\partial}{\partial|\textbf{r}'|}\langle \hat \varrho(\textbf{r}')\rangle \Bigg)
  \end{multline}
where $r_s=2GM/c^2$

With the definition of the matter stress-energy tensor \cite[p.164]{carroll2004introduction}, the variation of the matter interaction action (variation defined as $ \overline{\delta}$ to distinguish from $\delta$ for source of perturbation) produced  by the fluctuating term in the metric is \cite{adler2016gravitation}
\begin{eqnarray}
    \overline{\delta} S_{int}=-\frac{1}{2}\int d^4 x \sqrt{-g} T^{\mu\nu}h_{\mu\nu},
\end{eqnarray}
and the following matter interaction Hamiltonian is
\begin{equation}
    \overline{\delta}\hat{H}_{\rm int}=\hat{H}_{\rm int}=\frac{1}{2}\int d^3 r \sqrt{-g} h_{\mu\nu}\hat{T}^{\mu\nu}= \hat{H}^{(0)}_{\rm int}+ \hat{H}^{(1)}_{\rm int}.
\end{equation}
There is the well-known Hamiltonian for the Schr\"odinger-Newton equation in flat space-time \cite{bahrami2014schrodinger}:
\begin{eqnarray}
  \hat{H}^{(0)}_{\rm int}=\frac{c^2}{2} \int d^3 r h^{(0)}_{00}(\textbf{r}) \hat\varrho(\textbf{r})=-G  \int d^3 r\int  d^3 r' \frac{1}{|\textbf{r}-\textbf{r}'|}\langle \hat\varrho(\textbf{r}')\rangle\hat\varrho(\textbf{r}),
  \end{eqnarray}
and a found correction to Hamiltonian in weak external gravitational field:
    \begin{multline}
  \hat{H}^{(1)}_{\rm int}=\frac{c^2}{2} \int d^3 r \left(\frac{r_s}{|\textbf{r}|}h^{(0)}_{00}(\textbf{r}) \hat\varrho(\textbf{r})+q(\textbf{r}) \hat\varrho(\textbf{r})+\frac{(ct)^2}{2}\left(w(\textbf{r})\hat\varrho(\textbf{r})+h_{00}^{(0)}(\textbf{r})\delta^{ij}\phi(\textbf{r})_{,i} \hat\varrho(\textbf{r})_{,j}\right) \right)=\\=G r_s\int d^3 r\Bigg(\frac{1}{4\pi}\frac{1}{2}\int d^3 r' \int d^3 r'' \frac{1}{|\textbf{r}-\textbf{r}'||\textbf{r}'-\textbf{r}''||\textbf{r}''|^2}\frac{\partial}{\partial|\textbf{r}''|}\langle \hat \varrho(\textbf{r}'')\rangle\Bigg)\hat\varrho(\textbf{r})-\\-
  \frac{(ct)^2}{2}\int d^3 r\int  d^3 r' \left(\frac{1}{2|\textbf{r}-\textbf{r}'||\textbf{r}'|^2}\frac{\partial}{\partial|\textbf{r}'|}\langle \hat \varrho(\textbf{r}')\rangle\hat \varrho(\textbf{r})+\frac{1}{2|\textbf{r}-\textbf{r}'||\textbf{r}|^2}\langle \hat \varrho(\textbf{r}')\rangle\frac{\partial}{\partial|\textbf{r}|}\hat \varrho(\textbf{r})\right),
  \end{multline}
where we use transformation $\delta^{ij}\phi(\textbf{r})_{,i} \hat\varrho(\textbf{r})_{,j}=r_s/(2|\textbf{r}|^2)\partial_{|\textbf{r}|}\hat\varrho(\textbf{r})$.

Now, let's use Fourier transform as follow:
\begin{eqnarray}
  \hat\rho(\textbf{r})=\frac{1}{(2\pi)^3}\int d^3 k e^{i \textbf{k} \cdot (\hat{\textbf{r}}-\textbf{r})} \rho(\textbf{k})=\frac{1}{(2\pi)^3}\int d^3 k e^{-i \textbf{k}\textbf{r}} \hat L_\textbf{k}=\frac{1}{(2\pi)^3}\int d^3 k e^{i \textbf{k}\textbf{r}} \hat L^\dagger_\textbf{k},
  \end{eqnarray}
with $\hat L_\textbf{k}=e^{i \textbf{k}\hat{\textbf{r}}}\varrho(\textbf{k})$. The high-energy cut-off is implied in $\varrho(\textbf{k})$ \cite{bahrami2014schrodinger, nimmrichter2015stochastic}.

Then, after simple calculation we achieve:
\begin{gather}
    \int d^3 r\int  d^3 r' \frac{1}{|\textbf{r}-\textbf{r}'|}\langle \hat\varrho(\textbf{r}')\rangle\hat\varrho(\textbf{r})=\frac{1}{2\pi^2} \int d^3 k \frac{1}{k^2}\langle \hat{L}^\dagger_\textbf{k}\rangle \hat{L}_\textbf{k},\\
  \int d^3 r\int  d^3 r'\frac{1}{|\textbf{r}-\textbf{r}'||\textbf{r}'|^2}\frac{\partial}{\partial|\textbf{r}'|}\langle \hat \varrho(\textbf{r}')\rangle\hat \varrho(\textbf{r})=\frac{1}{4 \pi^4} \int d^3 k \int d^3 k' \frac{\textbf{k} \cdot (\textbf{k}'-\textbf{k})}{k'^2 |\textbf{k}-\textbf{k}'|^2}\langle \hat{L}^\dagger_\textbf{k}\rangle \hat{L}_{\textbf{k}'},\\
  \int d^3 r\int  d^3 r'\frac{1}{|\textbf{r}-\textbf{r}'||\textbf{r}|^2}\langle \hat \varrho(\textbf{r}')\rangle\frac{\partial}{\partial|\textbf{r}|}\hat \varrho(\textbf{r})=\frac{1}{4 \pi^4} \int d^3 k \int d^3 k' \frac{\textbf{k}' \cdot (\textbf{k}-\textbf{k}')}{k^2 |\textbf{k}-\textbf{k}'|^2}\langle \hat{L}^\dagger_\textbf{k}\rangle \hat{L}_{\textbf{k}'},\\
  \frac{1}{4\pi}\int d^3 r \int d^3 r' \int d^3 r'' \frac{1}{|\textbf{r}-\textbf{r}'||\textbf{r}'-\textbf{r}''||\textbf{r}''|^2}\frac{\partial}{\partial|\textbf{r}''|}\langle \hat \varrho(\textbf{r}'')\rangle\hat\varrho(\textbf{r})=\nonumber\\=\frac{1}{4 \pi^4} \int d^3 k \int d^3 k' \frac{\textbf{k} \cdot (\textbf{k}'-\textbf{k})}{k'^4 |\textbf{k}-\textbf{k}'|^2}\langle \hat{L}^\dagger_\textbf{k}\rangle \hat{L}_{\textbf{k}'}.
\end{gather}
The final expression for Hamiltonian $\hat{H}_{int}$ takes form:
\begin{eqnarray}
    \hat{H}_{int}=-\frac{G}{2\pi^2}\int d^3 k \int d^3 k' L_{\textbf{k}'} \langle L_\textbf{k}^\dagger \rangle F(\textbf{k}',\textbf{k}),\\
    F(\textbf{k}',\textbf{k})=\frac{1}{(2\pi)^3}\frac{\delta^3(\textbf{k}'-\textbf{k})}{|\textbf{k}'||\textbf{k}|}-\frac{r_s}{2\pi^2}\frac{1}{ |\textbf{k}-\textbf{k}'|^2}\Bigg(\frac{1}{2}\frac{\textbf{k} \cdot (\textbf{k}'-\textbf{k})}{k'^4}-\nonumber\\-\frac{(ct)^2}{2}\left(\frac{\textbf{k} \cdot (\textbf{k}'-\textbf{k})}{2\textbf{k}'^2}+\frac{\textbf{k}' \cdot (\textbf{k}-\textbf{k}')}{2\textbf{k}^2}\right)\Bigg).\label{F_function_app}
\end{eqnarray}

\section{Stochastic extension of the Schr\"odinger-Newton equation} \label{AppB}
The achieving Schr\"odinger-Newton equation is following:
\begin{equation}
    \dot{|\Psi\rangle}=-\frac{i}{\hbar}\left(\hat{H}_0+\hat{H}_{int}\right)|\Psi\rangle.
\end{equation}

A Markovian evolution equation for the state matrix $\rho$ of the system, known as \textit{a quantum master equation} in the most general form of the quantum master equation which is mathematically valid is the Lindblad form \cite{lindblad1976generators}:
\begin{equation} \label{markovian}
    \dot{\rho}=-\frac{i}{\hbar} [H,\rho]+c_k \rho c_k^\dagger-\frac{1}{2}\left\{c^\dagger_k c_k, \rho\right\}.
\end{equation}
Here $\left\{c_k \right\}$ is the ordered set of Lindblad operators \cite{wiseman2001complete}. We rewrite \eqref{markovian} for two Lindblad operators in form:
\begin{multline}
    \dot{\rho}=-\frac{i}{\hbar} [H,\rho]+\frac{G}{2\pi^2\hbar}\int d^3 k\int d^3 k' \alpha_{\textbf{k},\textbf{k}'}\left(A_\textbf{k} \rho A_{\textbf{k}'}^\dagger-\frac{1}{2}\left\{A_{\textbf{k}'}^\dagger A_{\textbf{k}}, \rho\right\}\right)+\\+i \frac{G}{2\pi^2\hbar}\int d^3 k\int d^3 k' \beta_{\textbf{k},\textbf{k}'}\left(B_\textbf{k} \rho B_{\textbf{k}'}^\dagger-\frac{1}{2}\left\{B_{\textbf{k}'}^\dagger B_{\textbf{k}}, \rho\right\}\right),
\end{multline}
with real kernels that have properties $\alpha_{\textbf{k},\textbf{k}'}=\alpha_{\textbf{k}',\textbf{k}}$ and $\beta_{\textbf{k},\textbf{k}'}=-\beta_{\textbf{k}',\textbf{k}}$ that are appeared from hermitivity of Lindblad operators. Because Lindblad equation \eqref{markovian} is invariant under c-number shifts as
\begin{eqnarray}
    c_k\longrightarrow c_k+\chi_k, \quad H\longrightarrow H+\frac{i \hbar}{2}(\chi_k^* c_k-\chi_k c_k^\dagger),
\end{eqnarray}
we can shift $A_k\longrightarrow A_k+a_k$ and $B_k\longrightarrow B_k+b_k$ and then get Hamiltonian:
\begin{eqnarray}
    H\longrightarrow H+\frac{i\hbar}{2}\int d^3 k\int d^3 k' \alpha_{\textbf{k},\textbf{k}'}\left(A_\textbf{k} a_{\textbf{k}'}^*-A_\textbf{k}^\dagger a_{\textbf{k}'}\right)-\frac{\hbar}{2}\int d^3 k\int d^3 k' \beta_{\textbf{k},\textbf{k}'}\left( B_\textbf{k} b_{\textbf{k}'}^*+B_\textbf{k}^\dagger b_{\textbf{k}'}\right).
\end{eqnarray}
We can compensate the non-linear term $\hat{H}_{int}$ by choosing suitable form of operators and shifts:
\begin{eqnarray}
    A_k=L_k,\\
    a_k=i\langle L_k\rangle,\\
    B_k=L_k,\\
    b_k=\langle L_k\rangle,
\end{eqnarray}
and get 
\begin{eqnarray}
    \alpha_{\textbf{k},\textbf{k}'}+\beta_{\textbf{k},\textbf{k}'}=F(\textbf{k}',\textbf{k}),
\end{eqnarray}
and find $\alpha_{\textbf{k},\textbf{k}'}$, $\beta_{\textbf{k},\textbf{k}'}$ by using the symmetric properties:
\begin{eqnarray}
    \alpha_{\textbf{k},\textbf{k}'}=\frac{F(\textbf{k}',\textbf{k})+F(\textbf{k},\textbf{k}')}{2},\\
    \beta_{\textbf{k},\textbf{k}'}=\frac{F(\textbf{k}',\textbf{k})-F(\textbf{k},\textbf{k}')}{2}.
\end{eqnarray}
So, the final master equation with modification, that balanced the non-linearity by Lindblad operator has form:
\begin{multline}\label{fulldensmat}
    \dot{\rho}=-\frac{i}{\hbar} [\hat{H}_0,\rho]+\frac{G}{2\pi^2\hbar}\int d^3 k\int d^3 k' \left(\frac{F(\textbf{k}',\textbf{k})+F(\textbf{k},\textbf{k}')}{2}+i\frac{F(\textbf{k}',\textbf{k})-F(\textbf{k},\textbf{k}')}{2}\right)\times\\ \times\left(L_\textbf{k} \rho L_{\textbf{k}'}^\dagger-\frac{1}{2}\left\{L_\textbf{k}^\dagger L_{\textbf{k}'}, \rho\right\}\right).
\end{multline}

\section{Calculation of integrals for decoherence rate estimation} \label{AppC}

It is necessary to make the Fourier transform of \eqref{F_function}:
\begin{equation}
    \int d^3 k\int d^3 k' F(\textbf{k},\textbf{k}') e^{i \textbf{k}\textbf{r}}e^{-i \textbf{k}'\textbf{r}'}= \frac{2\pi^2}{|\textbf{r}-\textbf{r}'|}-\frac{r_s}{2\pi^2}\left( \frac{1}{2}\int d^3 k\int d^3 k'\frac{\textbf{k} \cdot (\textbf{k}'-\textbf{k})}{k'^4}\frac{e^{i \textbf{k}\textbf{r}}e^{-i \textbf{k}'\textbf{r}'}}{ |\textbf{k}-\textbf{k}'|^2}\right),
\end{equation}
We have after cumbersome calculation:
\begin{multline}
\frac{1}{2}\int d^3 k\int d^3 k'\frac{e^{i \textbf{k}\textbf{r}}e^{-i \textbf{k}'\textbf{r}'}}{k'^4}\frac{\textbf{k} \cdot (\textbf{k}'-\textbf{k})}{ |\textbf{k}-\textbf{k}'|^2}=\\=\frac{1}{2}\int d^3 k\int d^3 k' \frac{e^{i \textbf{k}\textbf{r}}e^{-i \textbf{k}'\textbf{r}'}}{ k'^4} \left(\frac{1}{4\pi}\int d^3 x\frac{i \textbf{k}\cdot \textbf{x}}{x^3}e^{i (\textbf{k}-\textbf{k}')\textbf{x}}\right)=\\
=\frac{1}{2}\int d^3 k\int d^3 k' \frac{e^{i \textbf{k}\textbf{r}}e^{-i \textbf{k}'\textbf{r}'}}{ k'^2}\left(\frac{1}{4\pi}\int d^3 x'\frac{e^{i \textbf{k}'\textbf{x'}}}{x'}\right)  \left(\frac{1}{4\pi}\int d^3 x\frac{i \textbf{k}\cdot \textbf{x}}{x^3}e^{i (\textbf{k}-\textbf{k}')\textbf{x}}\right)=\\=
\frac{1}{2}\frac{1}{(4\pi)^2}\int d^3 k\int d^3 x\int d^3 x' e^{i \textbf{k}\textbf{r}}\frac{2\pi^2}{ x'|\textbf{r}'+\textbf{x}-\textbf{x}'|}\left(\frac{i \textbf{k}\cdot \textbf{x}}{x^3}e^{i \textbf{k}\textbf{x}}\right)=\\=\frac{1}{16}\int d^3 k\int d^3 x 2\pi\left(2A- |\textbf{r}'+\textbf{x}|\right)\left(\frac{i \textbf{k}\cdot \textbf{x}}{x^3}e^{i \textbf{k}(\textbf{x}+\textbf{r})}\right)=\\=
\frac{2\pi}{16}\int d^3 x \left(2A- |\textbf{r}'+\textbf{x}|\right)\left(\frac{1}{x^3}\textbf{x}\cdot \frac{\partial}{\partial \textbf{x}}(2\pi)^3 \delta^3(\textbf{r}+\textbf{x})\right)=\\=
-\pi^4\int d^3 x \delta^3(\textbf{r}+\textbf{x})  \frac{\partial}{\partial \textbf{x}}\cdot \left( \textbf{x}\left(2A- |\textbf{r}'+\textbf{x}|\right)\frac{1}{x^3}\right)=\pi^4 \frac{\textbf{r}\cdot(\textbf{r}-\textbf{r}')}{r^3|\textbf{r}-\textbf{r}'|},
\end{multline}
where $A$ is constant occurred in intermediate calculation.
So,
\begin{equation}
    \int d^3 k\int d^3 k' F(\textbf{k},\textbf{k}') e^{i \textbf{k}\textbf{r}}e^{-i \textbf{k}'\textbf{r}'}= \frac{2\pi^2}{|\textbf{r}-\textbf{r}'|}\left(1-\frac{r_s}{4}\frac{\textbf{r}\cdot(\textbf{r}-\textbf{r}')}{r^3}\right).
\end{equation}

\end{widetext}

\bibliography{base1}
\end{document}